\documentclass[epj]{webofc}
\usepackage[varg]{txfonts}   
\usepackage{lineno}
\begin{document}
%
%
\title{The Cherenkov Telescope Array}
\subtitle{Science Goals and Current Status}

\author{The CTA Consortium\inst{1}\fnsep\thanks{\email{info@cta-observatory.org}}, \and
        represented by Rene A. Ong\inst{2}\fnsep\thanks{\email{rene@astro.ucla.edu}} 
}

\institute{See \url{https://www.cta-observatory.org/consortium_authors/authors_2018_09.html} 
\and
        Department of Physics and Astronomy, University of California, Los Angeles CA, 90095, USA 
          }

\abstract{%
  The Cherenkov Telescope Array (CTA) is the major ground-based gamma-ray observatory planned for the next
  decade and beyond. Consisting of two large atmospheric Cherenkov telescope arrays (one in the southern hemisphere
  and one in the northern hemisphere), CTA will have superior angular resolution, a much wider energy range,
  and approximately an order of magnitude improvement in sensitivity, as compared to existing instruments.
  The CTA science programme will be rich and diverse, covering cosmic particle acceleration, 
  the astrophysics of extreme environments, and physics frontiers beyond the Standard Model.
  This paper outlines the science goals for CTA and covers the current status of the project.
}
\maketitle
\section{Introduction}

The field of very high-energy gamma-ray (VHE, E $> 20\,$GeV) astronomy has come of age
in recent years, due to a wealth of information from both ground-based
and space-borne telescopes.
Prior to 2000, the general expectation for the field was to detect the accelerators of
the high-energy cosmic rays, most likely supernova remnants (SNRs).
However, the reality over the last two decades has been the discovery of an astonishing
variety of VHE gamma-ray sources, both within the Milky Way,
e.g. SNRs, pulsars, pulsar wind nebulae (PWNe), cosmic ray
bombarded molecular clouds, stellar binaries,
massive stellar clusters, the supermassive black hole Sgr A*, etc.,
and outside, e.g. BL Lac objects, flat-spectrum radio quasars, radio galaxies,
starburst galaxies, Milky Way satellites, gamma-ray bursts (GRBs), etc., see \cite{TeVCAT}.
Most of these discoveries have been made by the major ground-based imaging atmospheric
Cherenkov telescopes (IACTs), H.E.S.S., MAGIC, and VERITAS, that have been joined in recent
years by the air shower array HAWC. 
Important contributions have also come from the space instruments Fermi-LAT and AGILE.

The recent results have been exciting and have established the field,
but there are indications that they represent the tip of the iceberg 
and that more source classes and many more sources remain to be discovered.
In addition, we know that the potential of the imaging atmopsheric Cherenkov technique
has not yet been fully exploited and that 
significant improvements in sensitivity and resolution over existing instruments can be achieved 
with an array of
Cherenkov telescopes that is both larger (to provide more views of the shower)
and more capable (e.g. with wider field-of-view and higher resolution cameras).
Thus, there is both strong scientific and technical motivation for a next-generation
observatory that further develops the imaging atmospheric Cherenkov technique.

\vspace{-0.25cm}

\begin{figure}[ht!]
\centering
\begin{minipage}{6.0cm}
\centering
\includegraphics[scale=0.41]{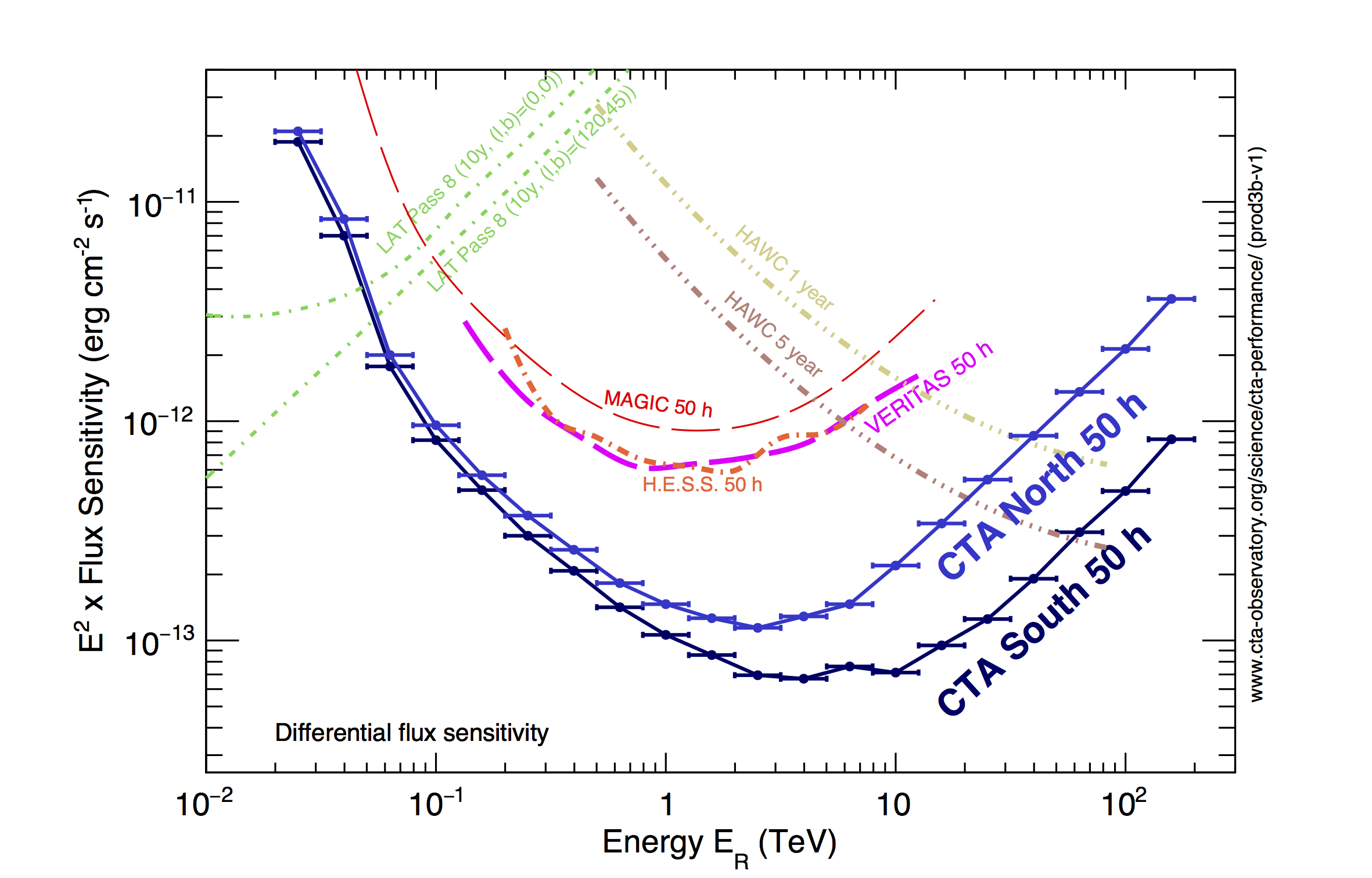}
\end{minipage}
\hspace{1.8cm}
\begin{minipage}{6.0cm}
\centering
\includegraphics[scale=0.33]{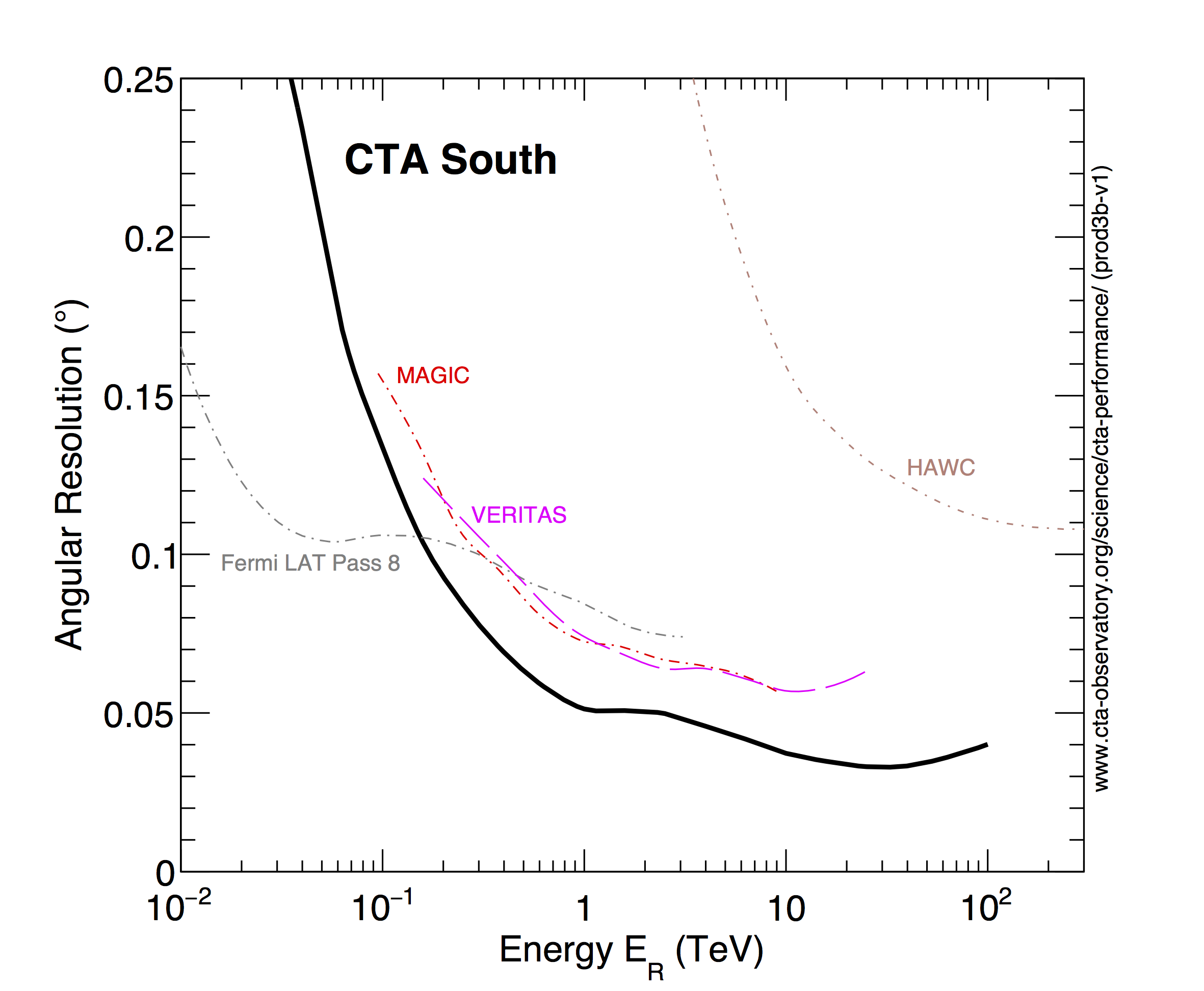}
\end{minipage}
\caption{{\bf Left:} Differential gamma-ray flux sensitivity (multiplied by E$^2$) for CTA-South
and CTA-North as a function of energy. 
Also shown are curves for the existing
IACTs (H.E.S.S., MAGIC and VERITAS), the air-shower experiment HAWC, and the Fermi-LAT
space telescope.  
{\bf Right:} Gamma-ray angular resolution for CTA-South as a function of energy, compared
to existing instruments.
For details on the sensitivity and angular resolution calculations and comparisons,
see \cite{CTA_Sensitivity}.
}
\label{fig-1}  
\end{figure}

\vspace{-0.25cm}

The Cherenkov Telescope Array (CTA) will transform our understanding of the VHE universe and
will address important questions in fundamental physics, such as the nature of dark matter.
It will also be an important member of the suite of experiments and observatories
participating in the expanding areas of multi-wavelength and multi-messenger astronomy.
Compared to the existing IACTs, CTA will: 1) cover a wider energy range, 2) have a significantly
larger field-of-view, and 3) achieve
up to an order of magnitude improvement in sensitivity (see Figure~\ref{fig-1}).
Angular resolution and energy resolution will also be improved and full-sky coverage will be ensured
by arrays in both the southern and northern hemispheres.
CTA will also be the first open observatory to operate in the VHE waveband, with approximately
50\% of the observing time set aside for guest observer proposals and all high-level data
available to the public after a proprietary period (typically one year).

CTA has been in development for more than a decade.
The concept was originated by the CTA Consortium, currently consisting of around 1,400 scientists
from 31 countries.
The Consortium has also developed the core science programme and has led the prototyping
of telescope hardware. 
The CTA Observatory (CTAO) has also been established; it is the legal entity that will oversee 
the construction and operation of the observatory.
Given a project the size and scope of CTA, international partnerships and good cooperation between 
the different stakeholders involved in the project (e.g. funding agencies, national laboratories, and scientific communities)
are required.

\section{CTA Key Science}

The core programme of science for CTA consists of major legacy studies,
called Key Science Projects (KSPs), to be carried out by the Consortium using guaranteed time
(comprising approximately 40\% of the observing time during the first ten-year period).
The KSPs have been developed and refined by Consortium members, 
with extensive input from scientists from outside of CTA.
The programme will be refined further as the observatory moves closer to its operational phase.
The currently proposed KSPs are described in an extensive document that
has been published as a book \cite{ScienceWithCTA}.

The core programme comprises three main scientific themes:
{\bf \em 1) Cosmic Particle Acceleration:}  How and where are particles accelerated, how do they
propagate and what is their impact on the environment?
{\bf \em 2) Probing Extreme Environments:} Understanding the physical processes close to neutron stars and
black holes,  characterizing relativistic jets, winds and explosions, and exploring radiation fields
and magnetic fields in cosmic voids.
{\bf \em 3) Physics Frontiers:} What is the nature of dark matter and how is it distributed?
Do axion-like particles exist? Is the speed of light a constant for high-energy photons?
These themes are developed fully within nine KSPs and the Dark Matter Programme.
Given space limitations, only highlights of a few KSPs are discussed here;
see \cite{ScienceWithCTA} for more complete information.

\subsection{Dark Matter Programme}

\begin{figure}[ht!]
\centering
\begin{minipage}{5.5cm}
\centering
\includegraphics[scale=0.33]{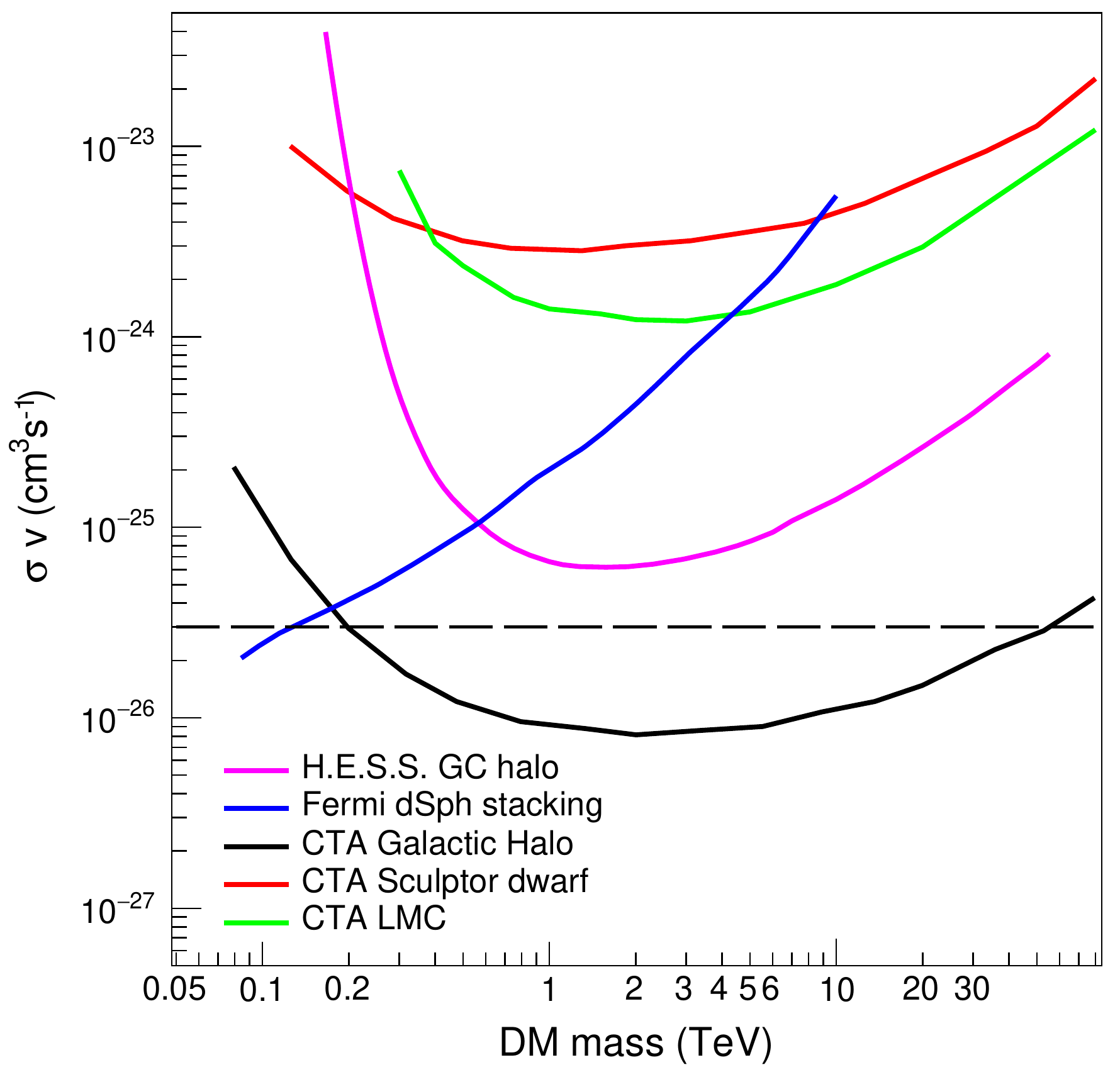}
\end{minipage}
\hspace{0.4cm}
\begin{minipage}{7.5cm}
\centering
\includegraphics[scale=0.55]{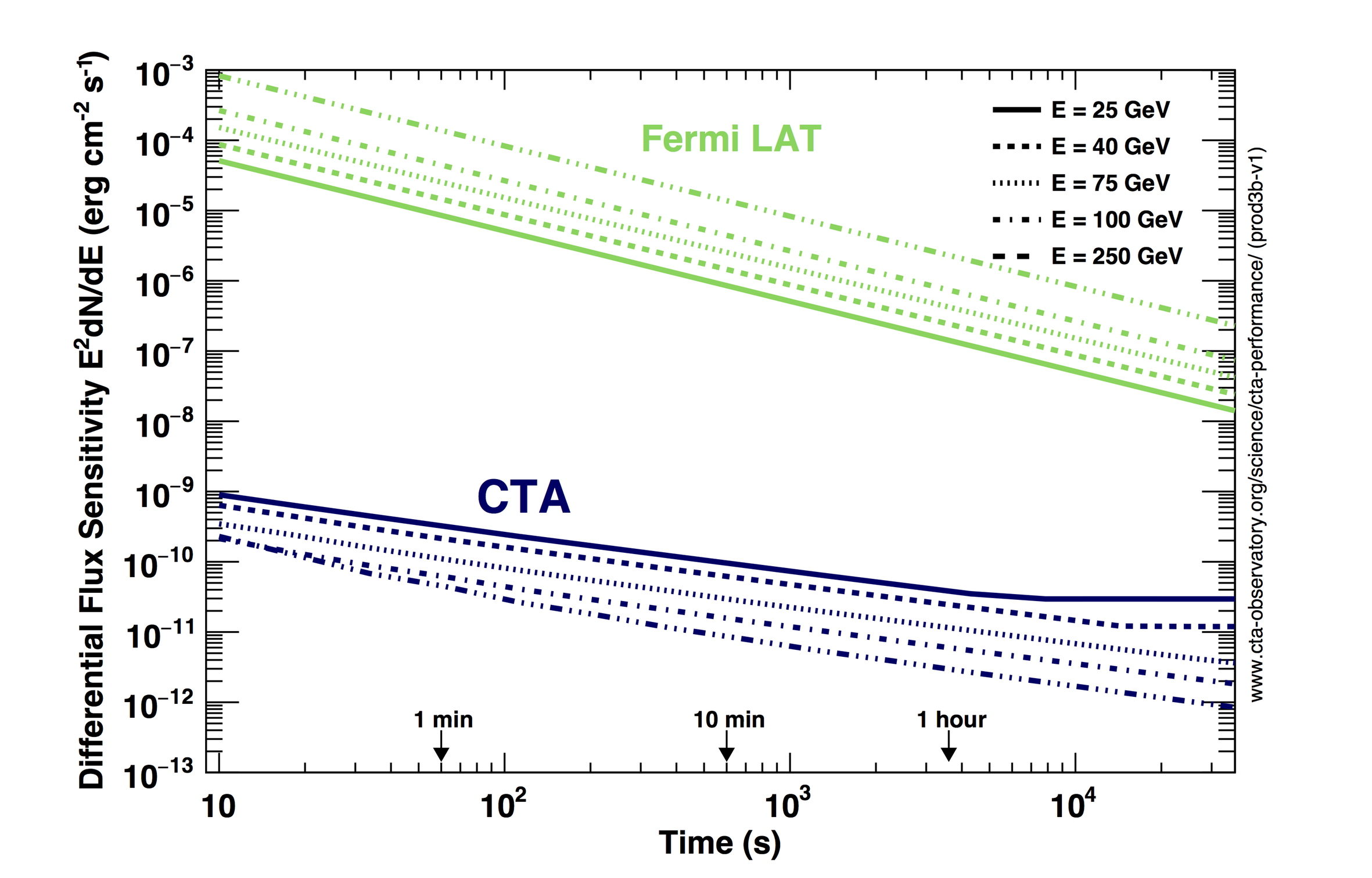}
\end{minipage}
\caption{{\bf Left:} 
Predicted CTA sensitivity on the velocity-weighted cross section
($\langle \sigma v \rangle$) for dark matter annihilation for various targets,
as indicated in the legend.
Results from H.E.S.S. on the Galactic centre halo \cite{HESS_DM} and from Fermi-LAT on
 dwarf spheroidal galaxies \cite{Fermi_DM} are also shown;  see
\cite{ScienceWithCTA} for details. 
{\bf Right:} Flux sensitivities for CTA and Fermi-LAT at selected energies as a function of observing
time; see \cite{CTA_Sensitivity} for details.
}
\label{fig-2}  
\end{figure}

The existence of dark matter is well established from a variety of astrophysical and cosmological
measurements.
CTA will search for dark matter via the indirect detection technique, looking for
gamma-ray signatures from the annihilation or decay of weakly interacting massive particles
(WIMPs).
The astrophysical targets for CTA are the Galactic centre halo, dwarf spheroidal galaxies, the
LMC and galaxy clusters.
The programme strategy focuses on maximizing the possibility of a detection.
Accordingly,
the primary target will be the Galactic centre halo that houses the largest concentration of dark matter
in our Galaxy, with secondary data taken on the other targets.
With a deep 500 h observation of the Galactic centre halo, CTA will achieve a sensitivity level below the benchmark 
relic cross-section over a wide WIMP mass range that extends into the multi-TeV region
(see Figure~\ref{fig-2}).
Thus, CTA will be an important player in the world-wide search for dark matter, providing unique and complementary information
to direct-detection experiments and the Large Hadron Collider.

\vspace{-0.05cm}

\vspace{-0.25cm}

\subsection{Transients KSP}
CTA is a powerful instrument for exploring the transient universe. With its very large collection area and
high signal to noise, CTA has outstanding instantaneous sensitivity over a broad energy range, giving
it a distinct advantage over Fermi-LAT at lower energies and HAWC at higher energies.
For transient time scales of a day or less, CTA is at least several orders of magnitude more sensitive than Fermi-LAT (see Figure~\ref{fig-2}).
CTA's disadvantage is that it has a limited field-of-view compared to Fermi-LAT or HAWC; 
hence, reaction time for CTA is critical and its
transients programme is focused on follow-up observations.
The Transients KSP is an integrated programme that encompasses a variety of
multi-wavelength and multi-messenger alerts, including explosive transients such as GRBs and
gravitational wave events, active galactic nuclei (AGN) flares, Galactic transients, and high-energy neutrino alerts.
Rapid dissemination to the wider community of the VHE properties of transients observed by CTA is an
important component of the KSP.

\subsection{Galactic Plane Survey KSP and Cosmic Ray PeVatrons KSP}

\begin{figure}[ht]
\centering
\includegraphics[scale=0.64]{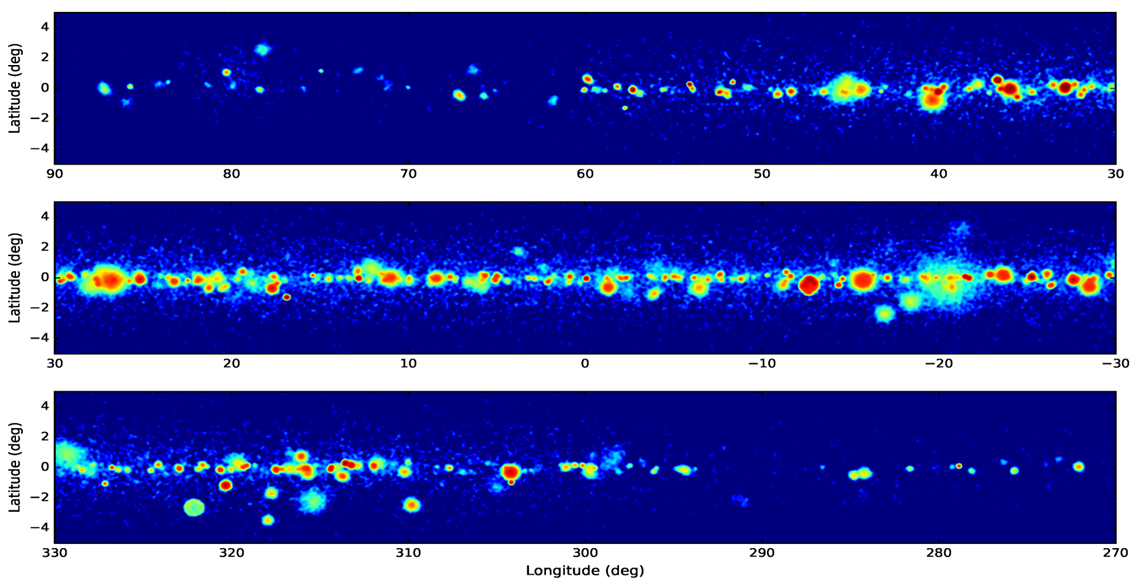}
\caption{Simulated results from the CTA Galactic Plane Survey in very high-energy gamma rays
for half of the plane.  The simulation
incorporates the proposed GPS observation strategy and a source model containing SNR and
PWN populations, as well as diffuse emission.}
\label{fig-3}      
\end{figure}

Surveys of the Galactic plane are key legacy programmes for all major observatories.
Previous surveys in the VHE 
gamma-ray band have been carried out by both ground-based and space-borne instruments.
The CTA Galactic Plane Survey (GPS) will be the first high sensitivity, high angular resolution survey
carried out at TeV energies. Using 1600 h of observing time, the GPS KSP will cover the entire Galactic plane
with a graded-exposure strategy, emphasizing key regions of the Galaxy in both the southern
and northern hemispheres. From simulations and our (incomplete) knowledge of the VHE source population, we
expect the GPS to discover many hundreds of new sources, especially PWNe, SNRs, and binaries.
The GPS will also have great potential for the discovery of new, unexpected phenomena and will 
provide a detailed view of diffuse VHE gamma-ray emission.
Figure~\ref{fig-3} shows a simulated image of the GPS results.

\vspace{-0.25cm}

A long-standing issue in high-energy astrophysics is to understand what sources
accelerate cosmic ray nuclei (hadrons) to the knee in the energy spectrum.
The standard paradigm involves diffusive shock acceleration in SNRs, but so far 
only a handful of SNRs provide good evidence for hadronic acceleration and
their energy spectra reach only to $\sim 10\,$TeV.
The goal of the Cosmic Ray Pevatrons KSP is to determine the sources capable of accelerating
hadrons to PeV energies.
For this, the GPS will be used as a finder and the
five brightest sources with spectral extension out to $\sim 100\,$TeV 
will be targeted for deep follow-up observations.
For any PeVatrons detected, correlation with data from other wavebands will be
crucial for clear identification of the physical sources.

\subsection{Extragalactic Survey KSP}

The motivation for a survey of the extragalactic sky is similar to that for the GPS;
we wish to observe a large portion of the sky in an uniform manner to get an understanding
of the dominant source population (in this case AGN) and to have sensitivity to completely
unexpected phenomena.
In the Extragalactic Survey KSP, one-quarter of the sky will be surveyed to a limiting sensitivity
of 0.6\% of the (standard candle) Crab nebula flux.
The survey will be carried out by both CTA-South and CTA-North and will connect to the GPS and cover the important
regions of Coma, Virgo, Cen A, and the northern Fermi Bubble.
The key science targeted by the survey includes the
unbiased determination of the luminosity function for AGN, the
discovery of VHE emission from as yet undetected source classes such as Seyfert galaxies and
ultraluminous infrared galaxies, and the detection of GRBs in the prompt phase.
Studies are ongoing to understand the potential of carrying out the survey using a divergent pointing strategy
(in which various telescopes are pointed to sky positions slightly offset from the target position), which would allow for larger instantaneous field-of-view.

\section{CTA Design and Current Status}

\subsection{CTA Design}

CTA will consist of two large arrays of imaging atmospheric Cherenkov telescopes.
The larger southern array, CTA-South, will be constructed in the Atacama Desert of Chile at an elevation
of 2,100\,m, close to the existing Paranal Observatory of the European Southern Observatory (ESO).
CTA-North, will be built at an elevation of 2,200\,m on the island of La Palma, Spain, located
within the existing Observatorio del Roque de los Muchachos of the Instituto de Astrofisica
de Canarias.
Two important administrative/scientific centers will be located in Europe:
the CTA Headquarters in Bologna, Italy and the CTA Science Data Management Center in Zeuthen, Germany.

\begin{figure}[ht!]
\centering
\begin{minipage}{5.8cm}
\centering
\includegraphics[scale=0.27]{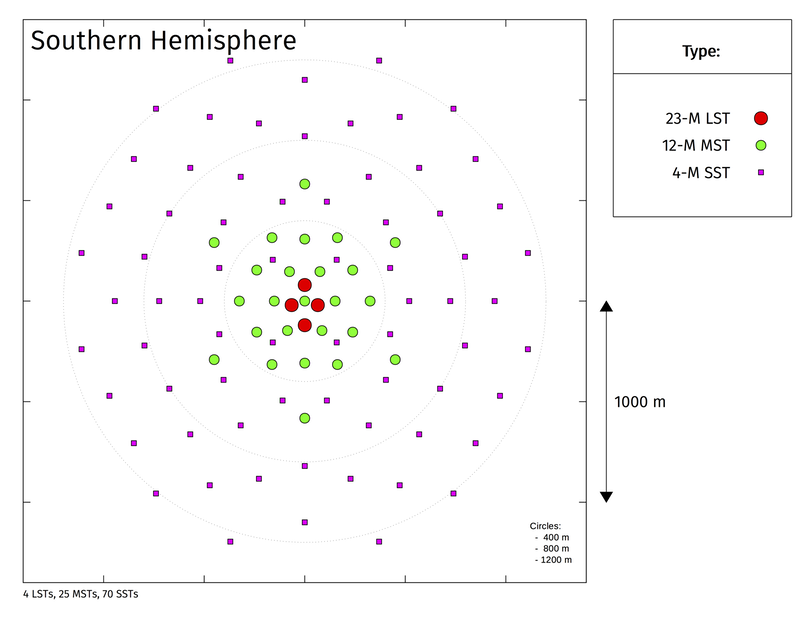}
\end{minipage}
\hspace{1.8cm}
\begin{minipage}{5.8cm}
\centering
\includegraphics[scale=0.30]{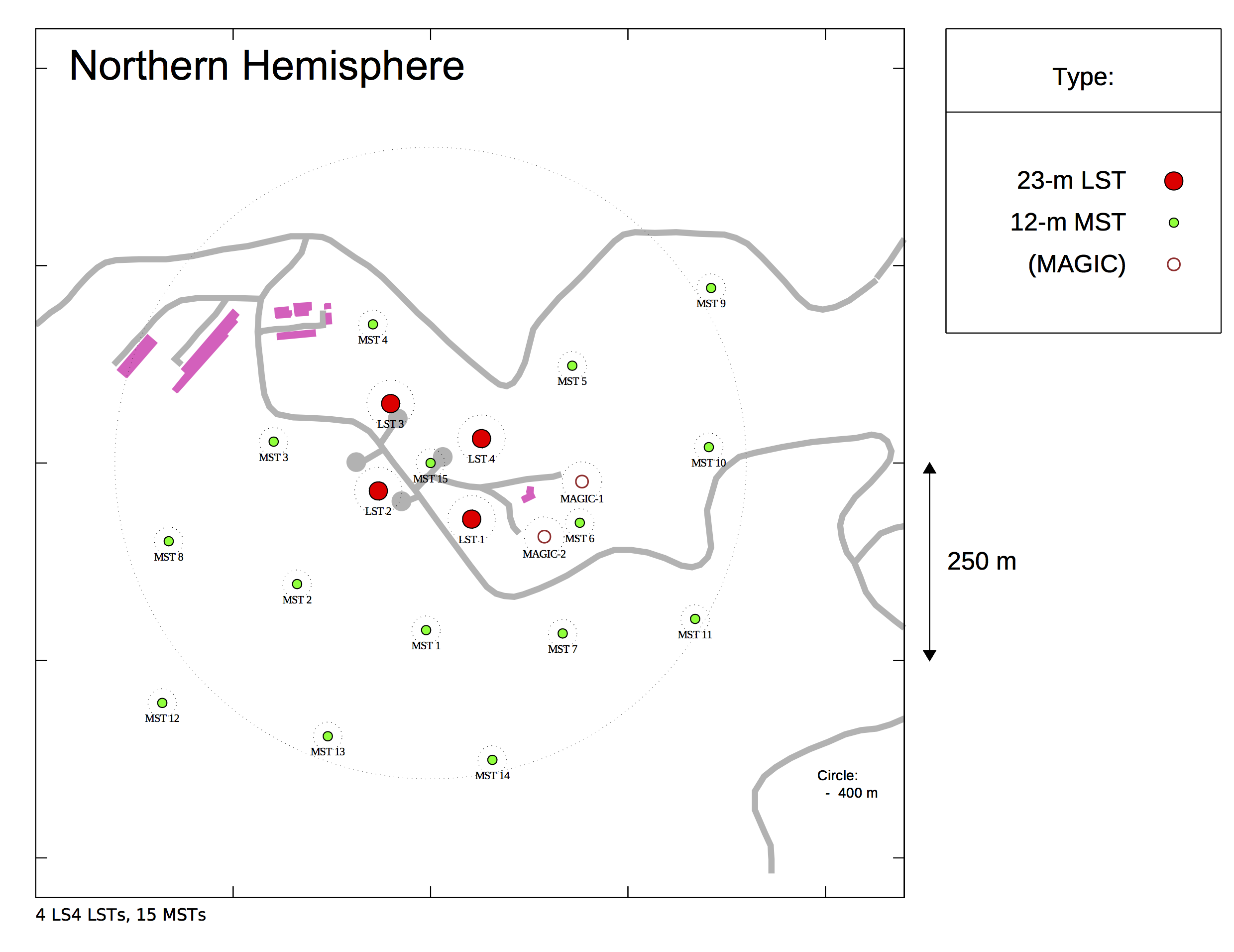}
\end{minipage}
\caption{{\bf Left:} 
Proposed layout for the three sizes of telescopes (LSTs, MSTs, and SSTs) in CTA-South (Paranal, Chile).
{\bf Right:} Proposed layout for the two sizes of telescopes (LSTs and MSTs) in CTA-North (La Palma, Spain).
The roads are shown in grey and the buildings in purple.  The current MAGIC telescopes are indicated by
the open red circles.
}
\label{fig-4}  
\end{figure}

In order to optimize the scientific capability within a constrained funding envelope, CTA will employ three different
sizes of telescopes \cite{CTA_Concept}.
Four Large-Sized Telescopes (LSTs), with mirror diameters of 23\,m,
will provide the sensitivity in the lowest energies of the CTA waveband (down to 20 GeV), where
the Cherenkov photon density in gamma-ray air showers is low ($\sim 1$/m$^2$).
Medium-Sized Telescopes (MSTs), with mirror diameters of 9-12\,m, will be distributed around the LSTs to provide sensitivity in the core energy range of 100\,GeV - 10\,TeV.
Small-Sized Telescopes (SSTs), with mirror diameters of $\sim 4\,$m, will extend the size of CTA-South 
and provide sensitivity to energies of 300\,TeV and above.
SSTs are only envisioned for the southern array for which the entire inner regions of the Galactic plane will be visible.
For both the MSTs and the SSTs, prototypes are being developed that use either
the conventional single-mirror Davies-Cotton optical design or the newly developed
dual-mirror Schwarzschild-Couder optical design.
In the baseline CTA design, CTA-South will comprise 4 LSTs, 25 MSTs, and 70 SSTs, covering an area of
$\sim 5\,$km$^2$, and CTA-North
will comprise 4 LSTs and 15 MSTs, covering an area of $\sim 0.5\,$km$^2$.
The proposed telescope array layouts, shown in Figure~\ref{fig-4},
were derived through a lengthy and comprehensive simulation process \cite{CTA_ArrayPaper}.

\subsection{CTA Status}

Initial planning for CTA started as early as 2005, when the community came together
at Palaiseau, France to discuss the future of ground-based gamma-ray astronomy
\cite{Palaiseau2005}.
Since then, an enormous amount of development work has been carried out and
CTA has been heavily reviewed and endorsed by the major roadmap committees
in particle physics, astroparticle physics, and astronomy 
(see, e.g., \cite{ASPERA,Astro2010,P5report}). 
Most recently, CTA was promoted to Landmark status
in the European Strategy Forum on Research Infrastructure (ESFRI) 2018 Roadmap
\cite{ESFRI2018}.
The management of CTAO has been established.
The CTA Council takes care of overall governance and the Project Office manages the 
construction project.
An interim legal entity (CTAO gGmbH), composed of shareholders from eleven countries and ESO, 
and associate members from
two countries, has been established for the preparation of the implementation of CTA.

\begin{figure}[ht]
\centering
\includegraphics[scale=0.39]{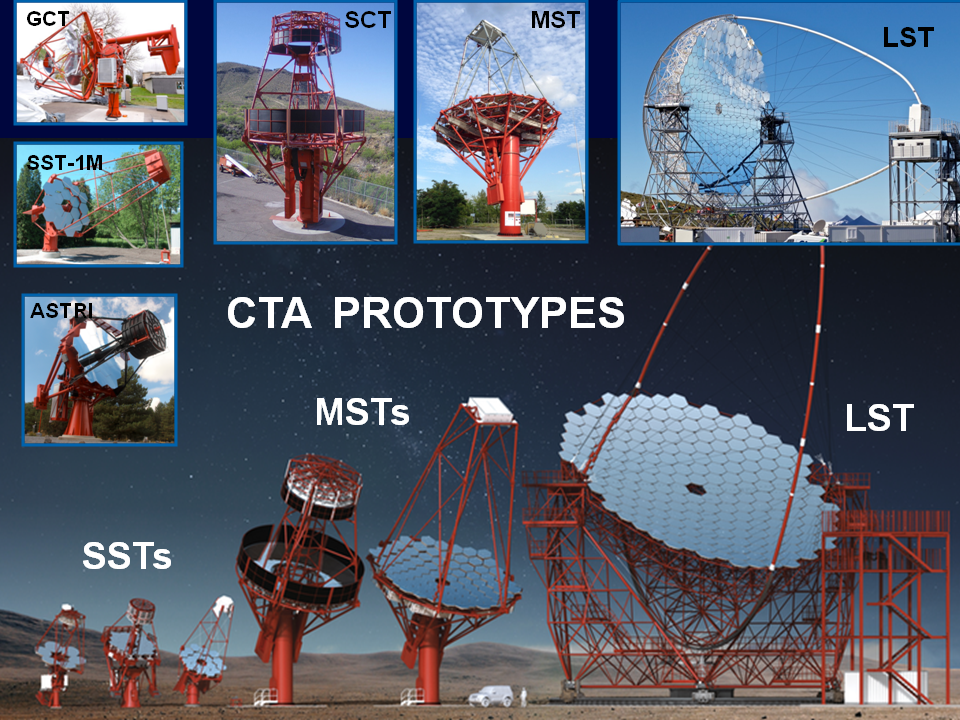}
\caption{Concept drawings and photographs of CTA protoype telescopes, as of 2018. Clockwise (from lower left):
SST-ASTRI at Mt. Etna, Italy, SST-1M at Krakow, Poland, SST-GCT at Meudon, France,
SCT at Whipple Observatory, AZ, USA, MST at Adlershof, Germany and LST-1 at La Palma, Spain.
}
\label{fig-5}      
\end{figure}

The development of telescope hardware is being carried out by the Consortium institutes,
as in-kind contributions to the observatory.
Substantial software development has also started, both at the Project Office and within Consortium institutes.
As shown in Figure~\ref{fig-5}, six full prototype telescopes exist and are being actively tested.
Work is now focused on design optimization, with the goal to reduce the number of telescope and camera options.
Site hosting agreements for both CTA sites have been signed 
and site preparatory work is ongoing. In this context, CTA-North is more advanced since much of the necessary infrastructure already
exists.
The first telescope to be located at a CTA site is the 
prototype LST that was inaugurated at the CTA-North site in October 2018 
and is now undergoing commissioning.
Current plans are: start of construction in 2019, first pre-production telescopes on site in 2021 (N) and 
2022 (S) and completion
of construction in 2025.
Since the CTA design is quite modular, early science operations can start as early as 2023, 
using a portion of the array available at either site.

\bibliography{bibliography_ong.bib}

\begin{thebibliography}{12}

\bibitem{TeVCAT}
TeVCAT: \url{http://tevcat.uchicago.edu/}

\bibitem{CTA_Sensitivity}
\url{https://www.cta-observatory.org/science/cta-performance/}

\bibitem{ScienceWithCTA}
{The Cherenkov Telescope Array Consortium}, \emph{{Science with the Cherenkov
  Telescope Array}} (World Scientific Publishing, 2019), ISBN
  {978-981-3270-08-4}, arXiv: 1709.07997, DOI: 10.1142/10986

\bibitem{HESS_DM}
H.~{Abdallah}, A.~{Abramowski}, F.~{Aharonian}, F.~{Ait Benkhali}, A.G.
  {Akhperjanian}, E.~{Ang{\"u}ner}, M.~{Arrieta}, P.~{Aubert}, M.~{Backes},
  A.~{Balzer} et~al., Physical Review Letters \textbf{117}, 111301 (2016),
  \texttt{1607.08142}

\bibitem{Fermi_DM}
M.~{Ackermann}, A.~{Albert}, B.~{Anderson}, W.B. {Atwood}, L.~{Baldini},
  G.~{Barbiellini}, D.~{Bastieri}, K.~{Bechtol}, R.~{Bellazzini}, E.~{Bissaldi}
  et~al., Physical Review Letters \textbf{115}, 231301 (2015),
  \texttt{1503.02641}

\bibitem{CTA_Concept}
B.S. {Acharya}, M.~{Actis}, T.~{Aghajani}, G.~{Agnetta}, J.~{Aguilar},
  F.~{Aharonian}, M.~{Ajello}, A.~{Akhperjanian}, M.~{Alcubierre},
  J.~{Aleksi{\'c}} et~al., Astroparticle Physics \textbf{43}, 3 (2013)

\bibitem{CTA_ArrayPaper}
A.~{Acharyya}, I.~{Agudo}, E.O. {Ang{\"u}ner}, R.~{Alfaro}, J.~{Alfaro},
  C.~{Alispach}, R.~{Aloisio}, R.~{Alves Batista}, J.P. {Amans}, L.~{Amati}
  et~al., Astroparticle Physics \textbf{111}, 35 (2019), \texttt{1904.01426}

\bibitem{Palaiseau2005}
B.~Degrange, G.~Fontaine, eds., \emph{{Proceedings, 7th Workshop on Towards a
  Network of Atmospheric Cherenkov Detectors 2005}}, Ecole Polytechnique (Ecole
  Polytechnique, Palaiseau, France, 2005)

\bibitem{ASPERA}
The AStroParticle ERAnet (ASPERA) roadmap: \url{http://www.aspera-eu.org/}

\bibitem{Astro2010}
The U.S. Astronomy and Astrophysics Decadal Survey (ASTRO2010):
  \url{http://www.nationalacademies.org/astro2010}

\bibitem{P5report}
The U.S. HEPAP P5 Panel: \url{http://usaparticlephysics.org/p5/}

\bibitem{ESFRI2018}
The ESFRI 2018 Roadmap: \url{http://roadmap2018.esfri.eu/}

\end{thebibliography}

\end{document}